\documentclass[trackchanges]{aastex701}

\usepackage{tikz}
\usetikzlibrary{calc} 
\begin{document}

\title{Observational Study of Multi-wavelength Synergistic Effects in 3C 120}

\author[0009-0007-2137-4641]{Yuting He}
\affil{State Key Laboratory of Radio Astronomy and Technology, Xinjiang Astronomical Observatory, Chinese Academy of Sciences, 150 Science 1-Street, Urumqi 830011, China}
\affil{School of Astronomy and Space Science, University of Chinese Academy of Sciences, No. 1 Yanqi East Road, Beijing 101408, China}
\email{heyuting@xao.ac.cn}

\author[0000-0002-8315-2848]{Ming Zhang}
\affil{State Key Laboratory of Radio Astronomy and Technology, Xinjiang Astronomical Observatory, Chinese Academy of Sciences, 150 Science 1-Street, Urumqi 830011, China}
\affil{School of Astronomy and Space Science, University of Chinese Academy of Sciences, No. 1 Yanqi East Road, Beijing 101408, China}
\email{zhangm@xao.ac.cn}
\correspondingauthor{zhangm@xao.ac.cn}

\author[0000-0003-4671-1740]{Qi Yuan}
\affil{Changchun Observatory, National Astronomical Observatories, Chinese Academy of Sciences, Changchun 130117, China}
\email{yuanqi@cho.ac.cn}

\author[0000-0001-6614-3344]{Nenghui Liao}
\affil{Department of Physics and Astronomy, College of Physics, Guizhou University, Guiyang 550025, China}
\email{nhliao@gzu.edu.cn}

\begin{abstract}
The energy dissipation and particle acceleration mechanisms within relativistic jets remain fundamental questions in active galactic nuclei (AGN) research. In this paper, we present a comprehensive 13-year (2012–2025) multi-wavelength study of the broad-line radio galaxy 3C 120, utilizing Fermi-LAT ($\gamma$-ray), ASAS-SN (optical), and high-resolution VLBA (15 GHz and 43 GHz) monitoring. Cross-correlation analyses reveal that $\gamma$-ray flares lead radio emission by $11.08_{-1.88}^{+4.03}$ months at 15 GHz and $8.27_{-5.55}^{+3.45}$ months at 43 GHz. This frequency-dependent temporal hierarchy positions the high-energy dissipation zone upstream of the radio core, corroborating the opacity-driven core-shift effect. By tracking the parsec-scale jet morphology during major $\gamma$-ray flaring epochs, we demonstrate that radio outbursts systematically coincide with compact core brightening, pronounced surges in polarized flux, abrupt electric vector position angle rotations, and the subsequent ejection of superluminal knots. Ultimately, our internal radio correlations suggest that jet dynamics are governed by a dual mechanism: long-term kinematic and flux baseline variations are geometrically modulated by a secular jet precession, while rapid, highly energetic polarimetric bursts are driven by short-lived internal shocks propagating down the jet channel.

\end{abstract}

\keywords{
  Active galactic nuclei (16);
Radio galaxies (1343);
Jets (870);
Gamma-ray astronomy (628);
Very long baseline interferometry (1769)
}
\section{Introduction} 
\label{sec:intro}

Relativistic jets launched by accreting supermassive black holes (SMBHs) in active galactic nuclei (AGN) are the most powerful persistent particle accelerators in the Universe, emitting non-thermal radiation from radio waves to TeV $\gamma$-rays \citep{Blandford1977, Zensus1997}. While the standard leptonic framework provides a successful description of the time-averaged broadband spectral energy distribution (SED) of radio-loud AGNs \citep{Ghisellini2010}, the dynamic processes linking high-energy dissipation to the morphological and magnetic evolution of the parsec-scale jet remain fundamental open questions. A critical challenge is deciphering how the initial energy release propagates downstream to alter local physical conditions during rapid flaring events \citep{Marscher2008, Tavecchio2021}. Recent multi-wavelength synergies address this by emphasizing the sequential causal connections between emission zones \citep{Kramarenko2022}. In the standard shock-in-jet scenario, upstream $\gamma$-ray flares---triggered by internal shocks or magnetic reconnection---propagate downstream, expanding and gradually becoming optically thin at radio frequencies. Observationally, this manifests as delayed radio outbursts, the emergence of superluminal knots, and complex electric vector position angle (EVPA) rotations \citep{Marscher2002, Marscher2008, Chatterjee2009, Liodakis2018}. Therefore, mapping the jet's opacity structure and unravelling its intrinsic drivers requires establishing a direct causal link between individual $\gamma$-ray outbursts and coincident structural or polarimetric changes. This task primarily relies on high-cadence, high-resolution very long baseline interferometry (VLBI) monitoring paired with continuous $\gamma$-ray coverage \citep{Casadio2015, Kramarenko2022}.

The broad-line radio galaxy 3C~120 ($z=0.033$) stands as one of the most luminous and active extragalactic jet sources in the local universe, offering a prime laboratory for probing the physics of high-energy dissipation in relativistic jets. As a nearby Fanaroff-Riley class I (FR I) source with a relativistic jet viewed at a moderate inclination angle ($\theta \sim 20^\circ$), 3C~120 enables detailed imaging of its parsec-scale jet structure and superluminal knot components, while avoiding the extreme Doppler beaming that often masks the intrinsic jet physics in highly aligned blazars \citep{Marscher2002, Chatterjee2009}. Its complex morphological and kinematic evolution has been widely discussed in the context of jet precession. Based on the periodicity in the historical optical light curves and the helical pattern observed in the large-scale jet structure, previous studies have provided strong evidence for a long-term jet precession in 3C~120 with a derived period of $\sim 12.3$ years \citep{Caproni2004}. This secular precession introduces periodic geometric viewing angle variations that modulate the observed jet emission across different bands. Furthermore, 3C~120 has been identified as a possible spatial and temporal counterpart to the high-energy neutrino event IC-180213A detected by the IceCube Neutrino Observatory \citep{2025A&A...702A.129C}. This association suggests that the jet of 3C~120 hosts efficient hadronic acceleration processes capable of producing PeV-scale particles, further underscoring the importance of characterizing its high-energy dissipation mechanisms and multi-wavelength flaring properties.

Previous statistical studies of large AGN samples have revealed a general temporal correlation between radio and $\gamma$-ray variability, typically finding that radio flares lag $\gamma$-ray outbursts by weeks to months, a delay commonly attributed to the frequency-dependent synchrotron opacity of the jet plasma and the light travel time between the $\gamma$-ray emission zone and the radio core \citep{Kramarenko2022}. Specifically for 3C~120, \citet{Casadio2015} and others have noted complex correlations between high-energy and radio variability, and identified a link between $\gamma$-ray flares and the ejection of new superluminal jet components \citep{Marscher2002, Chatterjee2009, Liodakis2018}. However, these prior works have been limited by either relatively short time baselines, sparse multi-frequency VLBI coverage, or a lack of detailed event-by-event analysis linking individual $\gamma$-ray flares to the concurrent evolution of jet kinematics, magnetic field topology, and spectral properties.

In this work, we investigate the multi-wavelength synergistic variability linked to high-energy flaring events in the broad-line radio galaxy 3C~120. Our primary goal is not only to evaluate the time delays between $\gamma$-ray outbursts and radio core emission at 15~GHz and 43~GHz, but also to systematically quantify the event-by-event co-evolution of jet kinematics and magnetic field topology. To achieve this, our observational strategy centers on tracing the downstream response of the radio jet to high-energy triggers. By defining the major $\gamma$-ray flaring epochs as temporal references, we investigate the correlated variability in radio flux density, magnetic field properties, and parsec-scale jet morphology, utilizing continuous $\gamma$-ray monitoring from \textit{Fermi}-LAT alongside 15~GHz (MOJAVE) and 43~GHz (BEAM-ME) VLBI data.

The paper is organized as follows: Section~\ref{sec:data} describes the data acquisition and processing of the multi-wavelength dataset. Section~\ref{sec:methods} details the cross-correlation methodologies. Section~\ref{sec:Results} presents the multi-wavelength correlation results, the derived radio flaring epochs with their corresponding polarization response characteristics, and the structural evolution of the parsec-scale jet. Section~\ref{sec:discussion} discusses the physical implications of our findings and the multi-wavelength kinematic connections. Finally, Section~\ref{sec:Conclusions} synthesizes the observed behaviors into a framework driven by long-term jet precession and transient internal perturbations.

\section{Data Acquisition and Processing}
\label{sec:data}
We have assembled a multi-wavelength dataset for 3C~120, covering a $\sim$13-year baseline from 2012 to 2025. In the following subsections, we detail the acquisition and reduction procedures for three observational domains: (i) high-resolution radio VLBI imaging at 15 and 43~GHz; (ii) optical photometry from ASAS-SN; and (iii) continuous $\gamma$-ray monitoring from \textit{Fermi}-LAT. 

\subsection{VLBI Data Processing and Model-fitting}
\label{subsec:vlbi_data}

The VLBI imaging dataset comprises observations from two major monitoring programs using the Very Long Baseline Array \citep[VLBA;][]{Napier1994}. The Ku-band data (centered at 15.4~GHz) were obtained from the MOJAVE program \citep{Lister2018} in dual-polarization mode, serving to investigate the structural evolution of the broader downstream jet. Complementary Q-band data (centered at 43.1~GHz) were retrieved from the VLBA-BU Blazar Monitoring Program \citep{Jorstad2017, Weaver2022}\footnote{BEAM-ME is the successor to the VLBA-BU-BLAZAR program, which conducts total and polarized intensity imaging of $\gamma$-ray-bright blazars at 43 and 86 GHz.} These observations, conducted in continuum mode with both left- and right-hand circular polarizations, were utilized to trace the innermost jet regions closest to the central engine.

\begin{figure*}[ht!]
\epsscale{0.8}
\plotone{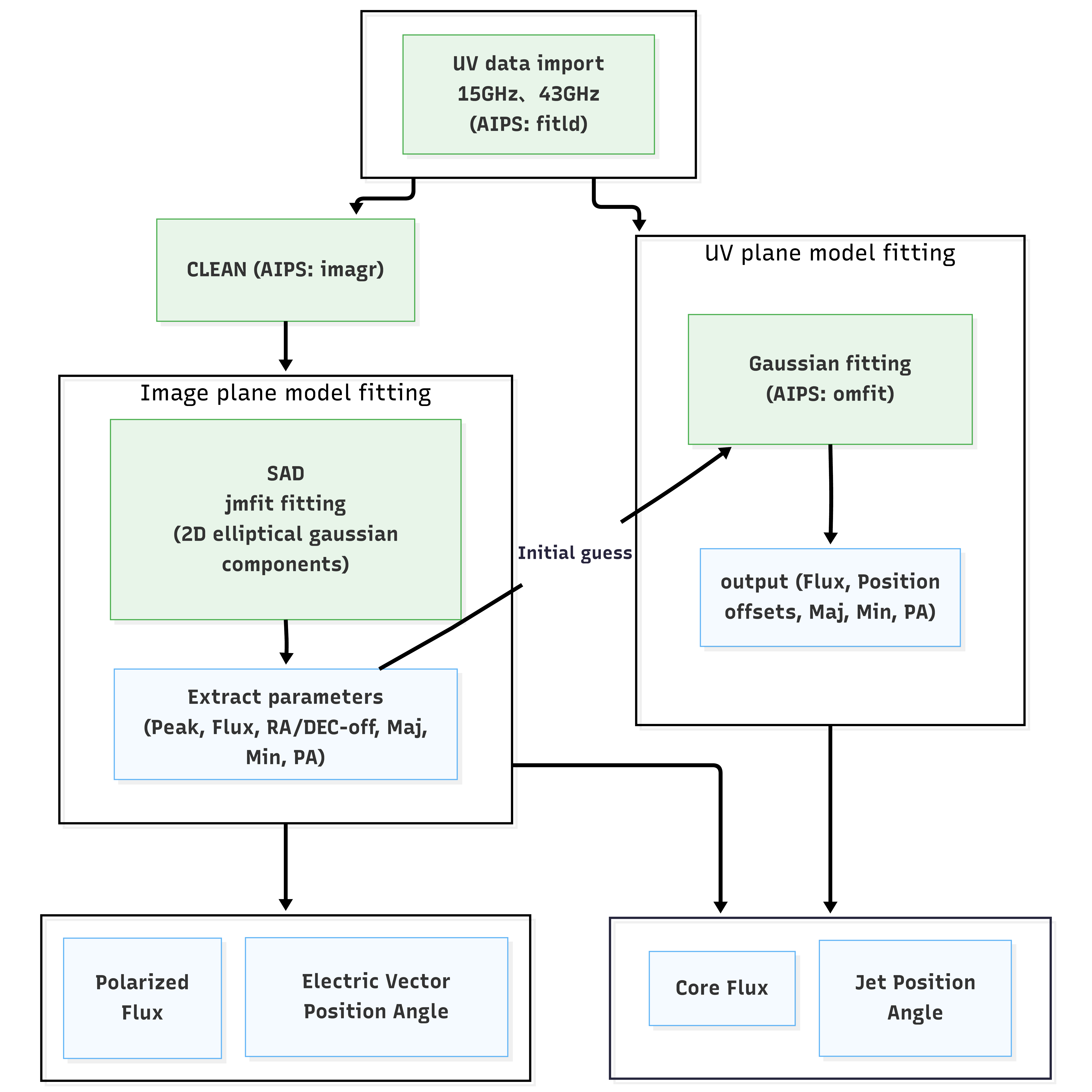}
\caption{Schematic of the SAND VLBI data-reduction pipeline, illustrating both $uv$-plane and image-plane fitting pathways.}
\label{fig:sand}
\end{figure*}

The VLBI data processing and parameter extraction were performed using the SAND pipeline \citep{Zhang2018}. While SAND is capable of deriving radio parameters in both the $uv$ and image planes, the physical observables utilized in this study were extracted from the image plane. This methodological choice is dictated by the morphological complexity of the parsec-scale jet in 3C 120. Analytical fitting in the $uv$ plane typically relies on idealized geometric priors, predominantly assuming a compact profile. In observational reality, however, the downstream jet knots of 3C 120 frequently evolve into irregular, non-Gaussian, or flat-topped structures. In contrast, the CLEAN algorithm non-parametrically reconstructs the true flux density distribution in the image plane, facilitating robust analysis of the complex morphology of fine structures. Image-plane model fitting is able to decompose the complex and extended structural profiles, regardless of their conformity to standard analytical shapes. Consequently, the image-plane model fitting provides a more robust and faithful representation of the jet's structural evolution. The complete data-reduction workflow schematic is illustrated in Figure~\ref{fig:sand}.

Building upon the image-plane model fitting, we extracted key structural and kinematic observables to trace the jet's evolution. First, the core flux density ($F_{\text{core}}$) is defined as the flux density of the fitted 2D elliptical Gaussian core component, which is identified as the brightest and most compact emission feature typically located at the upstream end of the jet. 

Second, the jet position angle (jetPA), which traces the macroscopic structural orientation of the downstream flow, is determined using the orthogonal distance regression (ODR) algorithm \citep{Yuan2023}. To trace the collimated flow, we employ individual resolved components within a defined core radius. We opted for this component-based approach to minimize the CLEAN-related systematics and multi-epoch inconsistencies often associated with continuous ridgeline procedures. For epochs with multiple resolved components, ODR derives the jetPA by performing a linear regression through their positional centroids, with the fitted line constrained to pass through the core. By minimizing the orthogonal distances from the data points to the best-fit ray, ODR provides robust fits across various slopes and simultaneously accounts for positional uncertainties in both right ascension and declination. During this regression, each component's coordinates are weighted by the inverse square of its positional uncertainty—a value derived directly from restoring beam and structural parameters such as the signal-to-noise ratio and deconvolved size. This weighting approach helps to reduce the influence of diffuse outer emission.

To characterize the magnetic field properties, the polarized flux density ($F_{\text{pol}}$) is derived from the combined Stokes $Q$ and $U$ maps ($P = \sqrt{Q^2 + U^2}$), representing the strength of the linear polarization. Finally, the electric vector position angle (EVPA) is derived as $\chi_{\text{pol}} = \frac{1}{2} \arctan(U/Q)$. To rectify the inherent $n\pi$ ambiguity, we applied an iterative correction algorithm following similar procedures in previous studies \citep[e.g.,][]{Larionov2016}: for consecutive epochs ($t_i, t_{i+1}$), $\chi_{\text{pol}}$ was adjusted by $\pm \pi$ whenever $|\Delta\chi_{\text{pol}}| > 2\pi/3$ ($120^\circ$). This minimizes unphysical jumps and preserves the continuity of the magnetic field orientation. A representative subset of these extracted results is presented in Table~\ref{tab:VLBI}.

\begin{deluxetable}{cccccc} 
\tabletypesize{\small}
\tablecaption{Key VLBI Observational Parameters of 3C 120 at 15 and 43 GHz\label{tab:VLBI}} 
\tablecolumns{6}
\setlength{\tabcolsep}{12pt}  
\tablehead{
  \colhead{Epoch} & \colhead{Frequency} & \colhead{$F_{\rm core}$} & \colhead{JetPA} & \colhead{EVPA} & \colhead{$F_{\rm pol}$} \\
  \colhead{} & \colhead{(GHz)} & \colhead{(Jy)} & \colhead{(deg)}  & \colhead{(deg)} & \colhead{(Jy)}
}
\startdata
2012-08-03 & 15 & $0.8309\pm0.0003$ & $-113.93\pm4.53$ & $-37.06\pm2.81$ & $0.0066\pm0.0017$ \\
2012-09-02 & 15 & $1.0331\pm0.0003$ & $-114.31\pm3.87$ & $-40.39\pm1.38$ & $0.0058\pm0.0015$ \\
2012-11-02 & 15 & $1.1627\pm0.0004$ & $-112.83\pm3.90$ & $-40.92\pm3.21$ & $0.0034\pm0.0009$ \\
2012-11-28 & 15 & $1.1794\pm0.0003$ & $-113.06\pm5.70$ & $-25.50\pm10.67$ & $0.0027\pm0.0006$ \\
2012-12-23 & 15 & $1.1493\pm0.0003$ & $-113.44\pm6.53$ & $-25.36\pm9.80$ & $0.0039\pm0.0012$ \\
2013-06-30 & 43 & $1.3071\pm0.0007$ & $-114.75\pm5.44$ & $83.76\pm1.68$ & $0.0029\pm0.0009$ \\
2013-07-29 & 43 & $0.6755\pm0.0006$ & $-116.21\pm5.84$ & $39.56\pm2.97$ & $0.0031\pm0.0013$ \\
2013-08-26 & 43 & $1.3506\pm0.0011$ & $-115.47\pm8.63$ & $127.45\pm2.38$ & $0.0038\pm0.0013$ \\
2013-11-18 & 43 & $1.0509\pm0.0007$ & $-116.05\pm4.85$ & $131.87\pm2.10$ & $0.0048\pm0.0018$ \\
2013-12-17 & 43 & $0.6331\pm0.0004$ & $-115.84\pm5.09$ & $136.84\pm0.25$ & $0.0017\pm0.0007$ \\
\enddata

\tablecomments{The observational data are VLBI results of 3C 120 at 15 GHz and 43 GHz; JetPA is the jet position angle, EVPA is the electric vector position angle, $F_{\rm core}$ is the core flux density. This table is available in its entirety in machine-readable form.
}
\end{deluxetable}

\subsection{Optical Photometry from ASAS-SN}
\label{subsec:optical_data}
We utilized the extensive optical monitoring data from the All-Sky Automated Survey for Supernovae (ASAS-SN; \citealt{Shappee2014, Kochanek2017, Hart2023}). The data were retrieved via the Sky Patrol interface\footnote{\url{http://asas-sn.ifa.hawaii.edu/skypatrol/}}, covering the period from 2012 to 2025. The observations comprise two distinct phases: $\text{v}$-band monitoring and $\text{g}$-band monitoring.

To construct a continuous optical light curve for 3C~120 spanning the period 2012--2025, we performed a zero-point alignment between the ASAS-SN $\text{v}$-band and $\text{g}$-band datasets. The $\text{v}$-band was adopted as the reference baseline. We calculated the systematic flux offset based on the mean flux levels of both bands and utilized this deviation to align the $\text{g}$-band measurements to the $\text{v}$-band system.

The calibrated $\text{g}$-band flux for the $j$-th observation, denoted as $F_{\text{g, corr}}^{(j)}$, is calculated as:
\begin{equation}
F_{\text{g, corr}}^{(j)} = F_{\text{g}}^{(j)} + \Delta F
\end{equation}
where $F_{\text{g}}^{(j)}$ represents the original observed flux density in the $\text{g}$-band. The systematic flux offset, $\Delta F$, is determined by the difference between the global mean fluxes of the two bands:
\begin{equation}
\Delta F = \bar{F}_{\text{v}} - \bar{F}_{\text{g}}
\end{equation}

Here, $\bar{F}_{\text{v}}$ and $\bar{F}_{\text{g}}$ correspond to the mean flux densities of the $\text{V}$ and $\text{g}$ bands, respectively. This alignment procedure ensures that the corrected $\text{g}$-band mean flux matches the $\text{v}$-band baseline ($\bar{F}'_{\text{g}} = \bar{F}_{\text{v}}$).

Although a simple additive offset cannot fully account for intrinsic AGN color evolution across different activity states, the v and g bands are sufficiently close in wavelength that they exhibit nearly identical short-term flaring morphologies. Because our primary objective is to evaluate macroscopic kinematic time lags rather than absolute photometry, preserving the temporal locations and relative variances of these rapid flares is sufficient.

Furthermore, standard cross-correlation algorithms are inherently sensitive to long-term baseline trends. Any uncorrected secular variations---such as steady accretion disk emission, host galaxy starlight, or residual slow color-induced drifts from the cross-filter transition---can bias the global normalization and dilute the temporal connections between different wavebands \citep{Chatterjee2009, LenTavares2010}. To simultaneously isolate the non-thermal jet kinematics and eliminate these slow baseline artifacts, we utilized a locally weighted scatterplot smoothing \citep[LOWESS;][]{Cleveland1979} algorithm, following established procedures in AGN light curve analyses \citep{Rakshit2020}. By adopting a temporal window fraction of 0.05, the LOWESS method models and subtracts the long-term secular baseline. This detrending process effectively purifies the fast-varying, non-thermal optical residuals, which subsequently serve as the refined input for our ZDCF analysis.

\subsection{Gamma-ray Monitoring from Fermi-LAT}
\label{subsec:gamma_data}

The $\gamma$-ray data utilized in this study are derived from the \textit{Fermi}-LAT observations processed by \citet{2025A&A...702A.129C}, who employed an adaptive-binning method to construct a light curve with uniform flux uncertainty, setting a target relative uncertainty of 20\% for each time interval. From this dataset, we adopted the median time of each adaptive bin as our observational epoch, paired with the corresponding photon flux and statistical uncertainty.

To identify physically motivated flaring epochs, \citet{2025A&A...702A.129C} applied the Bayesian Blocks algorithm, which systematically isolates extended periods of statistically significant flux elevation rather than symmetric temporal windows centered on instantaneous flux maxima. Based on this statistically robust segmentation, our analysis focuses on the three major $\gamma$-ray flaring epochs of 3C~120: Flare I (MJD~56861--56993), Flare II (MJD~57129--57298), and Flare III (MJD~58109--58298). Because high-energy flares in radio-loud AGNs frequently exhibit asymmetric temporal profiles---such as rapid flux surges followed by prolonged radiation cooling phases---the instantaneous flux maxima in our constructed light curve may not appear geometrically centered within these defined intervals. This apparent asymmetry is consistent with the intrinsic temporal dynamics of the jet's active states and our median-time binning approach, rather than a temporal misalignment.

It is worth noting that the $\gamma$-ray light curve exhibits additional flux enhancements. While these periods show clear variability, they do not satisfy the peak flux and duration thresholds established by the Bayesian Blocks analysis to be classified as major, independent flaring epochs. In this study, we classify these sub-threshold events as ``low-state $\gamma$-ray flares''.

\begin{figure*}[ht!]
\plotone{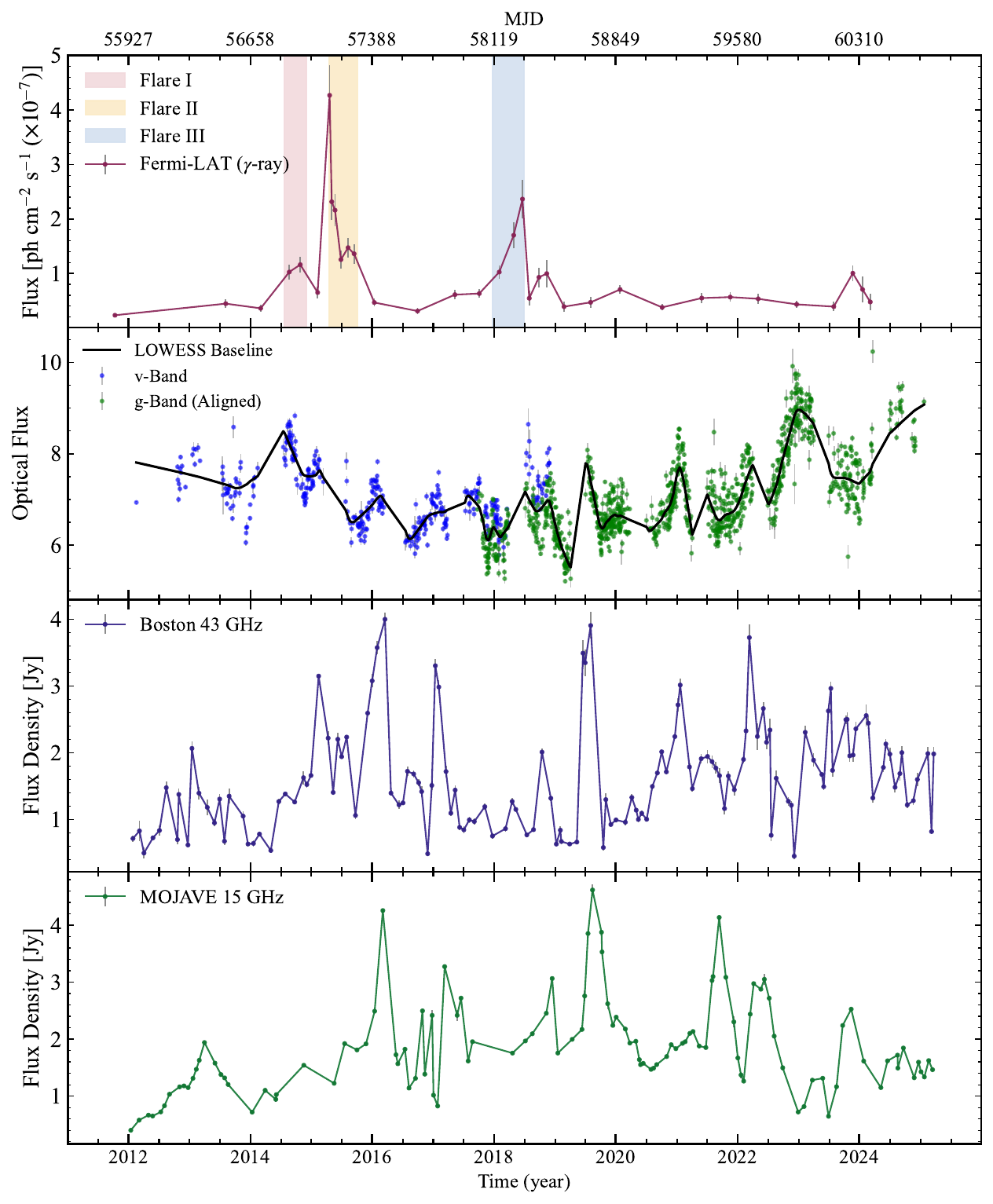}
\caption{Multi-wavelength light curves of 3C~120. The four vertically arranged panels display radiation variations across different bands: (top panel) $\gamma$-ray flux detected by Fermi-LAT; (second panel) aligned optical fluxes in the $v$ and $g$ bands from the ASAS-SN survey, superimposed with the long-term LOWESS baseline; (third panel) 43~GHz radio flux density from the VLBA-BU-BLAZAR monitoring program; (fourth panel) 15~GHz radio flux density from the MOJAVE program.}
\label{fig:1}
\end{figure*}

Figure~\ref{fig:1} presents the complete multi-wavelength light curves synthesized from these datasets, providing a comprehensive view of the long-term variability across the $\gamma$-ray, optical, 43 GHz, and 15 GHz bands.

\section{Analysis Methods}
\label{sec:methods}
The cross-correlation function (CCF) quantifies the temporal correlation between two time series, $x(t)$ (e.g., $\gamma$-ray flux) and $y(t)$ (e.g., radio flux density), as a function of the time lag $\tau$. The normalized CCF is defined as:

\begin{equation}
\text{CCF}(\tau) = \frac{1}{N_\tau \sigma_x \sigma_y} \sum_{i} \left[ x(t_i) - \bar{x} \right] \left[ y(t_i + \tau) - \bar{y} \right]
\label{eq:ccf_definition}
\end{equation}
where $\tau$ denotes the time lag between the two series; $N_\tau$ is the number of valid data pairs used for the CCF calculation at lag $\tau$; $x(t_i)$ and $y(t_i)$ are the observed values at the $i$-th epoch $t_i$; $\bar{x}$ and $\bar{y}$ are the arithmetic means, and $\sigma_x$ and $\sigma_y$ are the standard deviations of the $x(t)$ and $y(t)$ series, respectively. A prominent peak in the CCF profile at lag $\tau_0$ indicates that variability in $y(t)$ lags behind that in $x(t)$ by a characteristic time $\tau_0$, which is typically interpreted as reflecting the physical propagation or radiative response timescale of emission within an AGN jet.

However, standard discrete CCF methods assume uniformly sampled data. Given the uneven sampling and seasonal gaps inherent in astronomical monitoring \citep{Edelson1988}, we employ two complementary variants: the interpolation cross-correlation function \citep[ICCF;][]{Gaskell1986, White1994} and the $z$-transformed discrete cross-correlation function \citep[ZDCF;][]{Alexander1997}.

\subsection{Multi-wavelength Cross-Correlation (ZDCF)}
\label{subsec:method_zdcf}

\begin{figure*}[ht!]
\epsscale{1.1}
\plotone{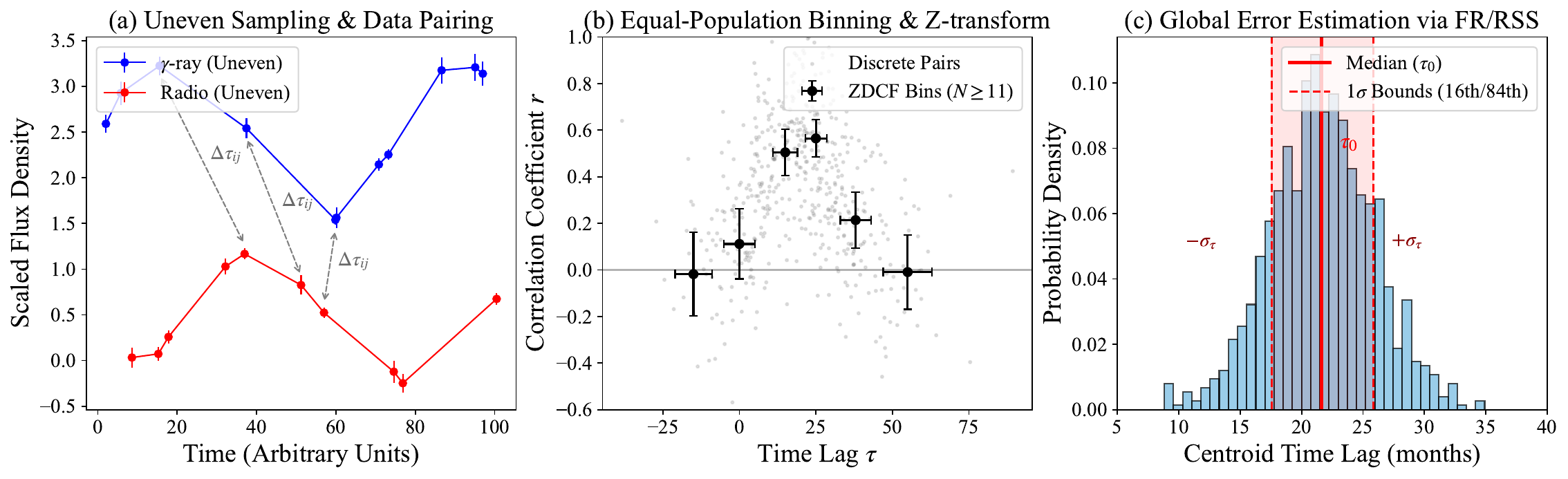}
\caption{Schematic of the ZDCF methodology and global error estimation. \textbf{(a)} Data pairing and discrete time lag ($\Delta\tau_{ij}$) extraction from unevenly sampled light curves. \textbf{(b)} Equal-population binning ($N_{min}=11$) of discrete pairs to construct the ZDCF profile. \textbf{(c)} Centroid distribution from monte carlo FR/RSS simulations, defining the final physical time lag (median) and its $1\sigma$ uncertainties. Detailed definitions of the statistical parameters and error bounds are provided in Section \ref{subsec:method_zdcf}.
\label{fig:ZDCF}}
\end{figure*}

To quantify the temporal correlation and physical time lags between the unevenly sampled Fermi-LAT $\gamma$-ray data and the radio light curves, we employ the ZDCF(\citep{Alexander1997}). The ZDCF algorithm is specifically optimized for sparse, unevenly sampled astronomical time series and avoids the systematic biases often introduced by interpolation-based methods, as illustrated in Figure~\ref{fig:ZDCF}a. Its robustness for investigating $\gamma$-ray and radio temporal connections has been extensively validated in recent large-scale AGN studies \citep{Kramarenko2022}. Following standard practice, we adopt an equal-population binning strategy with a minimum of $N_{\text{min}}=11$ independent data pairs per bin to ensure uniform statistical significance across the correlation profile. A critical aspect of our methodology is the quantification of uncertainties.

For each individual time-lag bin, we report two sets of error bounds, as shown in Figure~\ref{fig:ZDCF}b. The horizontal error bars represent the temporal width of the bin, defined by the minimum and maximum time lags of the data pairs contained within each bin. The vertical error bars denote the $1\sigma$ statistical uncertainty of the correlation coefficient, which is analytically derived using the Fisher z-transform method to correct for finite sample size biases \citep{Alexander1997}. To propagate this uncertainty to the global time-lag measurement of the correlation peak, we implement a monte carlo approach combining flux randomization (FR) and random subset selection (RSS). We generate 1,000 synthetic light curve pairs. In each iteration, the observation epochs are randomly resampled with replacement, and the corresponding flux measurements are perturbed by gaussian noise scaled to their respective observational uncertainties. For each simulated light curve pair, we compute the cross-correlation profile and extract the centroid of the peak, defined as the flux-weighted mean lag of all bins with correlation coefficients exceeding 80\% of the maximum value. The final reported time lag is taken as the median of the resulting centroid distribution, with its $1\sigma$ confidence interval defined by the 16th and 84th percentiles of the 1,000 simulations, as demonstrated in Figure~\ref{fig:ZDCF}c.

\subsection{Internal Radio Cross-Correlation (ICCF)}
\label{subsec:method_ccf}

\begin{figure*}[ht!]
\epsscale{1.1}
\plotone{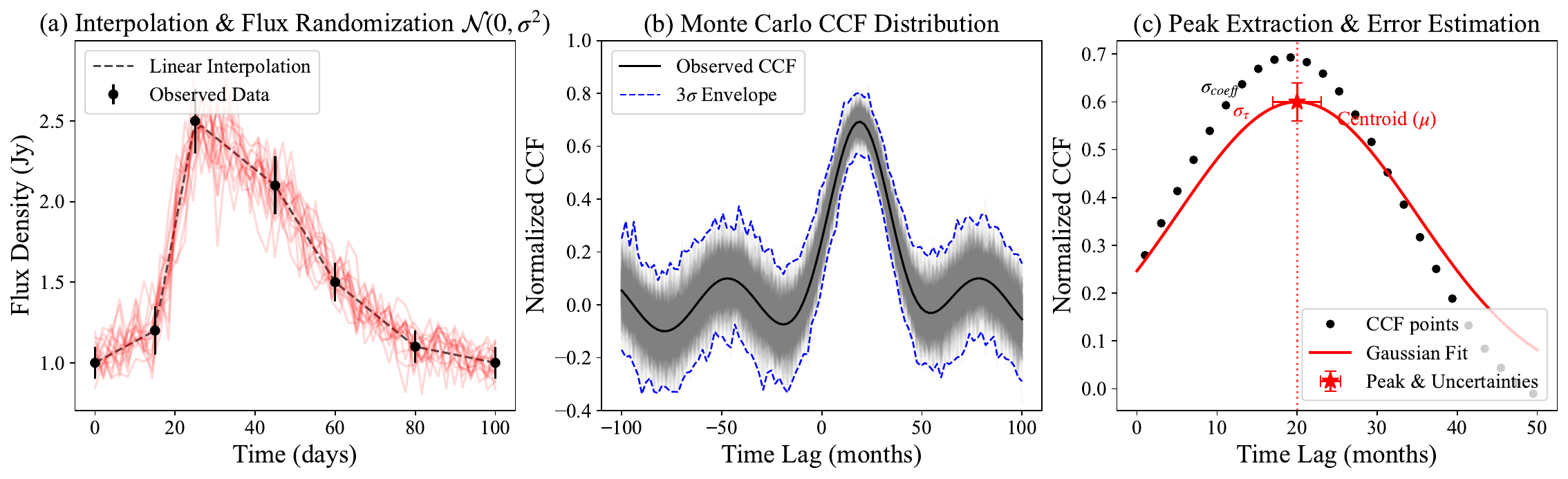}
\caption{Schematic of the ICCF methodology and error estimation. \textbf{(a)} Linear interpolation and flux randomization of observed data. \textbf{(b)} Simulated CCF ensemble defining the $3\sigma$ confidence envelopes. \textbf{(c)} Peak centroid ($\mu$) and uncertainties ($\sigma_{\tau}$, $\sigma_{coeff}$) extracted via gaussian fitting and monte carlo statistics. Detailed definitions are provided in Section \ref{subsec:method_ccf}.
\label{fig:ICCF}}
\end{figure*}

To investigate the internal temporal couplings among the multi-frequency radio jet observables---specifically core flux density, polarized flux density, jet position angle, and electric vector position angle---we utilize the ICCF. A schematic overview of this analytical workflow is presented in Figure~\ref{fig:ICCF}.

Unlike the sparsely sampled $\gamma$-ray data, our VLBA monitoring datasets have sufficiently high sampling density to support robust linear interpolation. While the ZDCF was necessary in Section~\ref{subsec:method_zdcf} to avoid artificial interpolation biases, its equal-population binning scheme makes it inherently suboptimal here. Applying ZDCF to our dense radio data would artificially widen the time-lag bins, smearing the cross-correlation profile and degrading the temporal resolution required to resolve subtle, short-timescale delays. Therefore, the ICCF is better suited for tracing these fine-scale internal couplings. By resampling the irregularly spaced radio light curves onto a fine, uniform temporal grid, the ICCF generates a continuous cross-correlation profile. To safely preserve intrinsic high-frequency variability without introducing artificial smoothing, the interpolation step size was empirically set to one-third of the median sampling interval of the original observations (Figure~\ref{fig:ICCF}a). This continuous profile is critical, as it enables the extraction of highly precise, sub-epoch time lags.

Because standard cross-correlation calculations do not intrinsically propagate measurement errors, we coupled the ICCF with a robust monte carlo flux randomization framework to evaluate statistical significance. We performed $N_{\text{sim}} = 10,000$ independent simulations. In each iteration, the interpolated fluxes for both parameters were perturbed by injecting Gaussian noise scaled to their corresponding interpolated measurement uncertainties, effectively sampling from $\mathcal{N}(0, \sigma^2)$ for each epoch. The CCF was recalculated for each simulated pair to construct a probability density distribution of the correlation coefficients at every lag (Figure~\ref{fig:ICCF}b). From this ensemble, we extracted the 0.15\% and 99.85\% percentiles to construct a $3\sigma$ (99.73\%) confidence envelope, explicitly visualizing how observational flux uncertainties propagate into the correlation strength.

Finally, to determine the characteristic physical time lag, we fitted a gaussian profile to the local neighborhood of the primary correlation peak within this envelope (Figure~\ref{fig:ICCF}c). The reported physical time lag corresponds to the centroid ($\mu$) of the best-fit gaussian, with its $1\sigma$ uncertainty ($\sigma_\tau$) extracted directly from the covariance matrix of the fitting routine. Concurrently, the statistical uncertainty of the maximum correlation coefficient is derived from the standard deviation of the monte carlo CCF distribution at the identified peak lag. This comprehensive methodology effectively accounts for the morphological width of the correlation peak, providing mathematically robust constraints on the physical delays between the jet's kinematic evolution and polarimetric outbursts.

\section{Results}
\label{sec:Results}

\subsection{Multi-wavelength Correlations and Flare Analysis}
\label{subsec:Correlations}
\begin{figure*}[ht!]
\epsscale{1.1}
\plotone{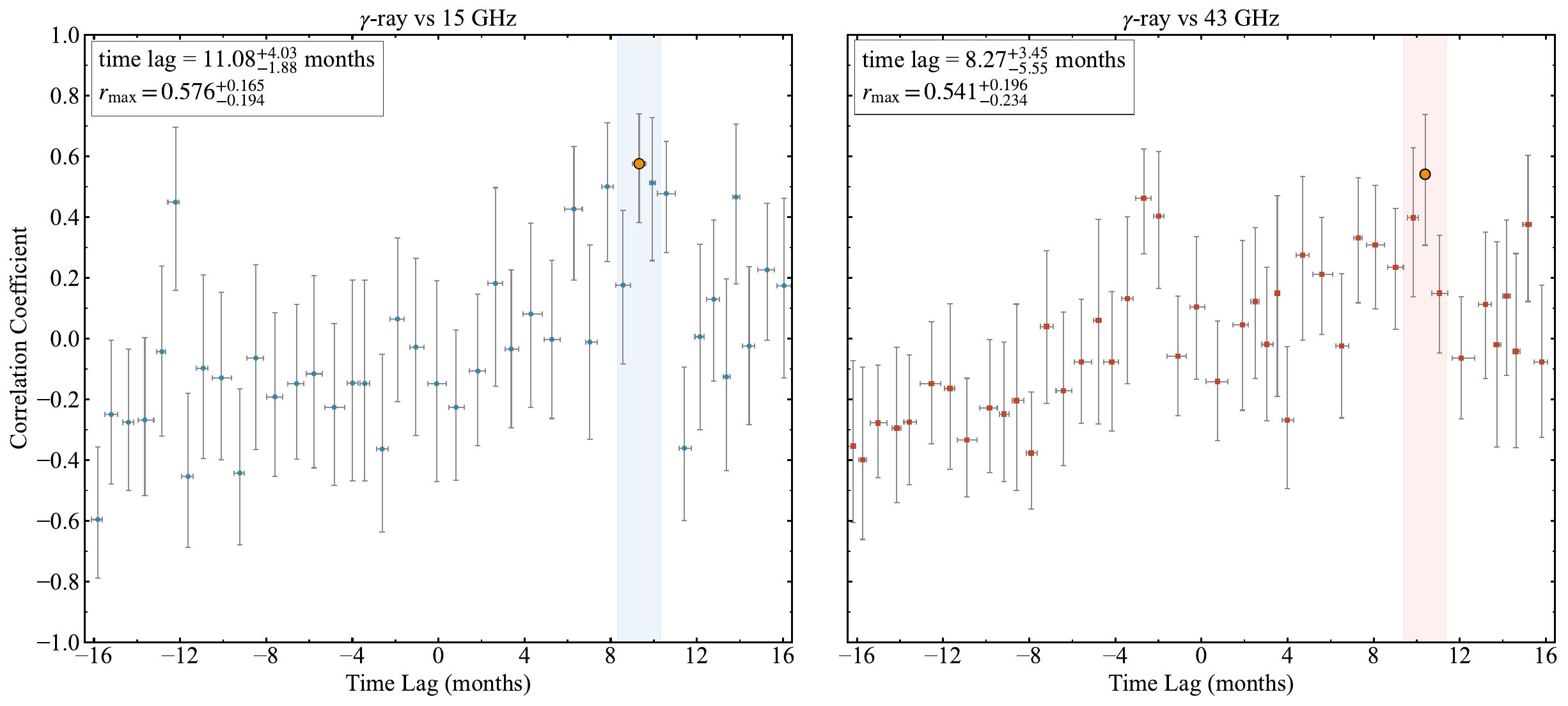}
\caption{ZDCF profiles for the cross-correlation between \textit{Fermi}-LAT $\gamma$-ray flux and radio core flux densities of 3C 120. 
\label{fig:2}}
\end{figure*}

We performed a cross-correlation analysis between the \textit{Fermi}-LAT $\gamma$-ray flux and the radio core flux densities at 15~GHz and 43~GHz using the ZDCF. The resulting profiles (Figure~\ref{fig:2}) reveal significant positive correlations, demonstrating that the high-energy $\gamma$-ray variations lead the radio emission. Specifically, the $\gamma$-ray/15~GHz correlation peaks at $r = 0.576$ with a time lag of $\tau_{\text{15GHz}} = 11.08_{-1.88}^{+4.03}$~months, while the 43~GHz correlation peaks at $r = 0.541$ with a shorter lag of $\tau_{\text{43GHz}} = 8.27_{-5.55}^{+3.45}$~months. These positive lags ($\tau > 0$) indicate that the $\gamma$-ray dissipation zone is spatially located upstream of the radio cores. We note that when their $1\sigma$ uncertainties are taken into account, these two inferred time lags overlap and are statistically consistent with each other.

The cross-correlation profiles also exhibit secondary peaks at negative time lags. We attribute these features to mathematical artifacts likely arising from the inherent red-noise nature of AGN light curves, as cross-correlating finite, stochastic time series naturally induces spurious aliasing \citep{Max-Moerbeck2014}. More critically, any negative time lag implies that radio flares precede $\gamma$-ray outbursts---a scenario that is inconsistent with the standard jet opacity paradigm. Upstream high-energy dissipation regions are extremely dense and highly optically thick to synchrotron self-absorption at radio frequencies. Radio emission only becomes observable later, after the emitting region propagates downstream, expands, and becomes optically thin \citep{Pushkarev2010, Fuhrmann2014}. Therefore, to avoid misinterpreting red-noise artifacts as physical kinematics, we restrict our analysis to the primary positive correlation peaks. Furthermore, the correlation analysis we conduct in Section 4.3, which includes multi-dimensional observables of flux, jet PA, EVPA and morphological structure, can further substantiate the effectiveness of our multi-band light curve correlation from another perspective.

\begin{figure*}[ht!]
\epsscale{1.1}
\plotone{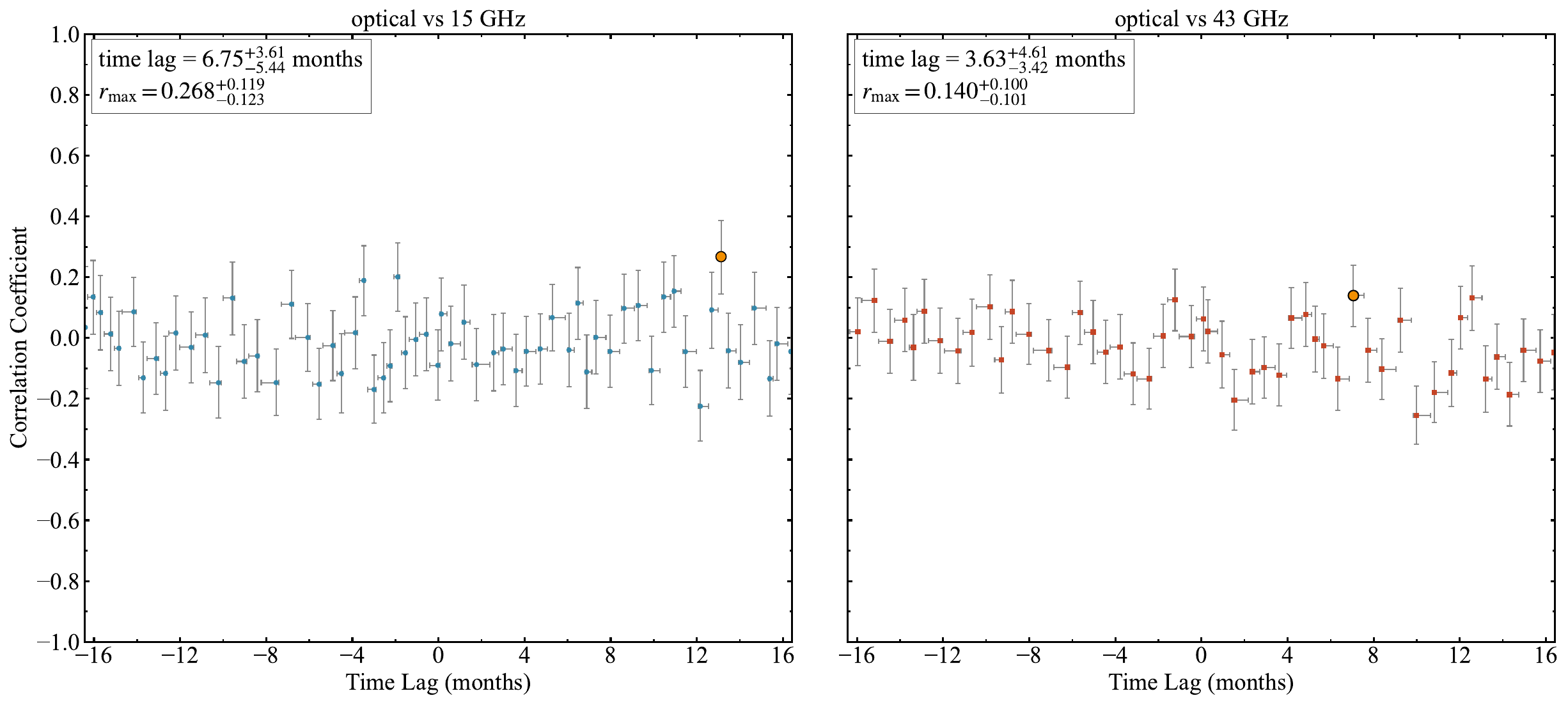}
\caption{ZDCF profiles for the cross-correlation between the LOWESS-detrended optical residuals and radio core flux densities of 3C~120.
\label{fig:guang}}
\end{figure*}

In addition to the high-energy $\gamma$-ray correlations, we performed a parallel ZDCF analysis utilizing the LOWESS-detrended non-thermal optical residuals to probe the optical-radio kinematic connection. Unlike the prominent $\gamma$-ray correlations, this analysis yielded relatively flat ZDCF profiles, with maximum correlation coefficients ($r_{\text{max}} \approx 0.14$--$0.27$) in Figure~\ref{fig:guang}.

\subsection{Radio Flaring Epochs and Polarization Responses}
\label{subsec:radio_flaring_polarization}

\begin{figure*}[ht!]
\epsscale{1.2}
\plotone{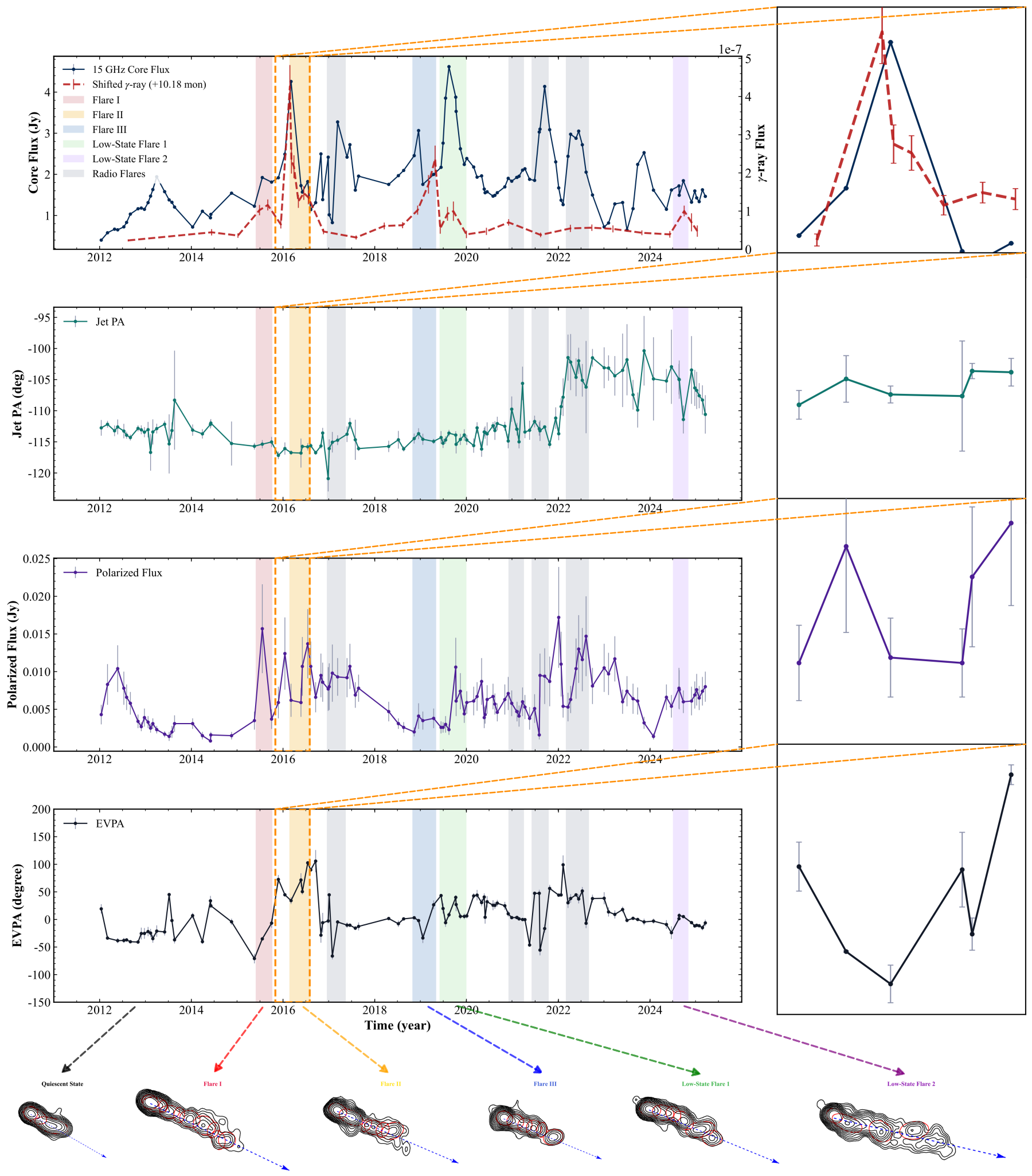}
\caption{Multi-parameter light curves and corresponding parsec-scale morphological evolution of 3C~120 at 15~GHz from the MOJAVE monitoring program. Top panels: From top to bottom, the light curves display core flux density, jet position angle, polarized flux, and electric vector position angle. The right panels provide a magnified view of Flare II. Bottom panels: Sequence of 15~GHz VLBA total intensity contour maps comparing the parsec-scale jet structure during the quiescent baseline and across the respective active flaring intervals, with the morphology of each flaring interval illustrated by a representative epoch from its corresponding observing window.}
\label{fig:3}
\end{figure*}

\begin{figure*}[ht!]
\epsscale{1.2}
\plotone{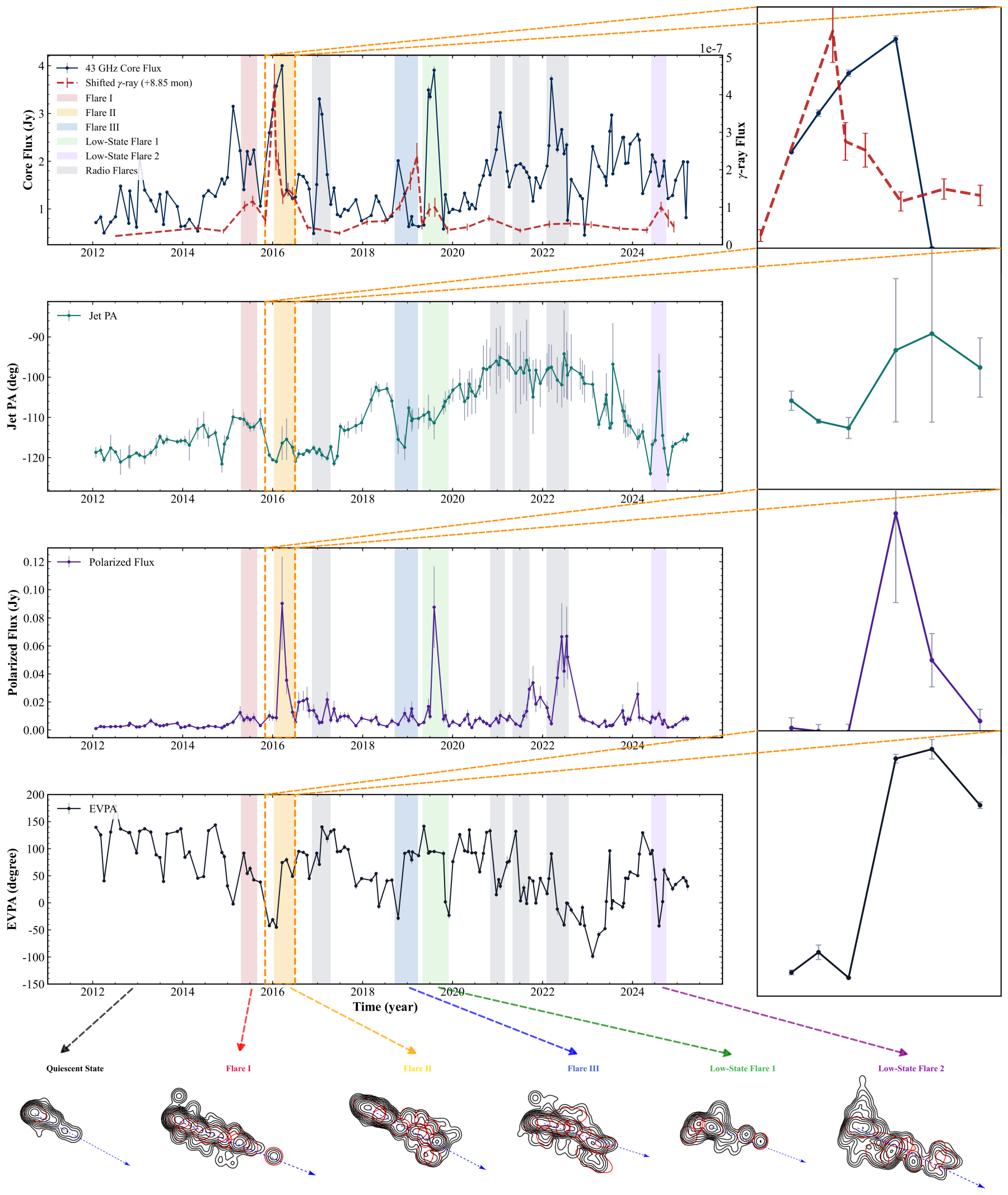}
\caption{Multi-parameter light curves and corresponding parsec-scale morphological evolution of 3C~120 at 43~GHz from the VLBA-BU-BLAZAR monitoring program. Top panels: From top to bottom, the light curves display core flux density, jet position angle, polarized flux, and electric vector position angle. The right panels provide a magnified view of Flare II. Bottom panels: Sequence of 43 GHz VLBA total intensity contour maps comparing the parsec-scale jet structure during the quiescent baseline and across the respective active flaring intervals, with the morphology of each flaring interval illustrated by a representative epoch from its corresponding observing window.}
\label{fig:4}
\end{figure*}

By superimposing the \textit{Fermi}-LAT $\gamma$-ray light curve---shifted by $+11.08$~months at 15~GHz and $+8.27$~months at 43~GHz---directly onto the radio core time series, we provide an intuitive visual representation of the temporal coupling between the high-energy and radio emission bands. As presented in Figures~\ref{fig:3} and \ref{fig:4}, the shifted $\gamma$-ray peaks show good alignment with the delayed radio flaring windows, demonstrating consistent multi-wavelength temporal correspondence.

The three major $\gamma$-ray flaring epochs identified in Section~\ref{subsec:gamma_data} (Flare I, II, and III) and their corresponding radio responses are highlighted by the red, yellow, and blue shaded regions, respectively. In addition to these prominent outbursts, the two low-state $\gamma$-ray flares discussed marked by the green and purple shaded regions also display clear radio counterparts. Crucially, the superimposed light curves demonstrate that even these weaker high-energy fluctuations are followed by corresponding delayed structural and flux responses in the downstream radio jet.

Across all these $\gamma$-ray-associated flaring epochs, we observe a physical coupling between the total core flux density and its polarimetric properties. Specifically, outbursts in the radio core flux are systematically accompanied by significant surges in polarized intensity and concurrent abrupt rotations of the electric vector position angle, indicating dramatic restructuring of the local magnetic field topology. To explicitly illustrate these synchronized fine-scale dynamics, we provide magnified inset panels focusing on Flare II in both Figures~\ref{fig:3} and \ref{fig:4}. As detailed in these insets, the 15~GHz and 43~GHz core flux densities exhibited dramatic rises from 2.49~Jy to 4.25~Jy and 2.59~Jy to 4.0~Jy, respectively. Coinciding closely with these total flux peaks, the polarized flux experienced several-fold enhancements (from 0.005~Jy to 0.013~Jy at 15~GHz, and 0.01~Jy to 0.09~Jy at 43~GHz), accompanied by rapid EVPA swings.

Finally, the gray shaded regions in Figures~\ref{fig:3} and \ref{fig:4} denote specific radio flaring epochs. These radio flares exhibit polarimetric behaviors consistent with those of the canonical synergistic events. They are similarly characterized by abrupt variations in total core flux density, substantial surges in polarized emission, and sharp EVPA rotations.

\subsection{Internal Radio Correlations and the Dual-mechanism Scenario}
\label{subsec:internal_radio_correlations}

\begin{figure*}[tbp!]
\plotone{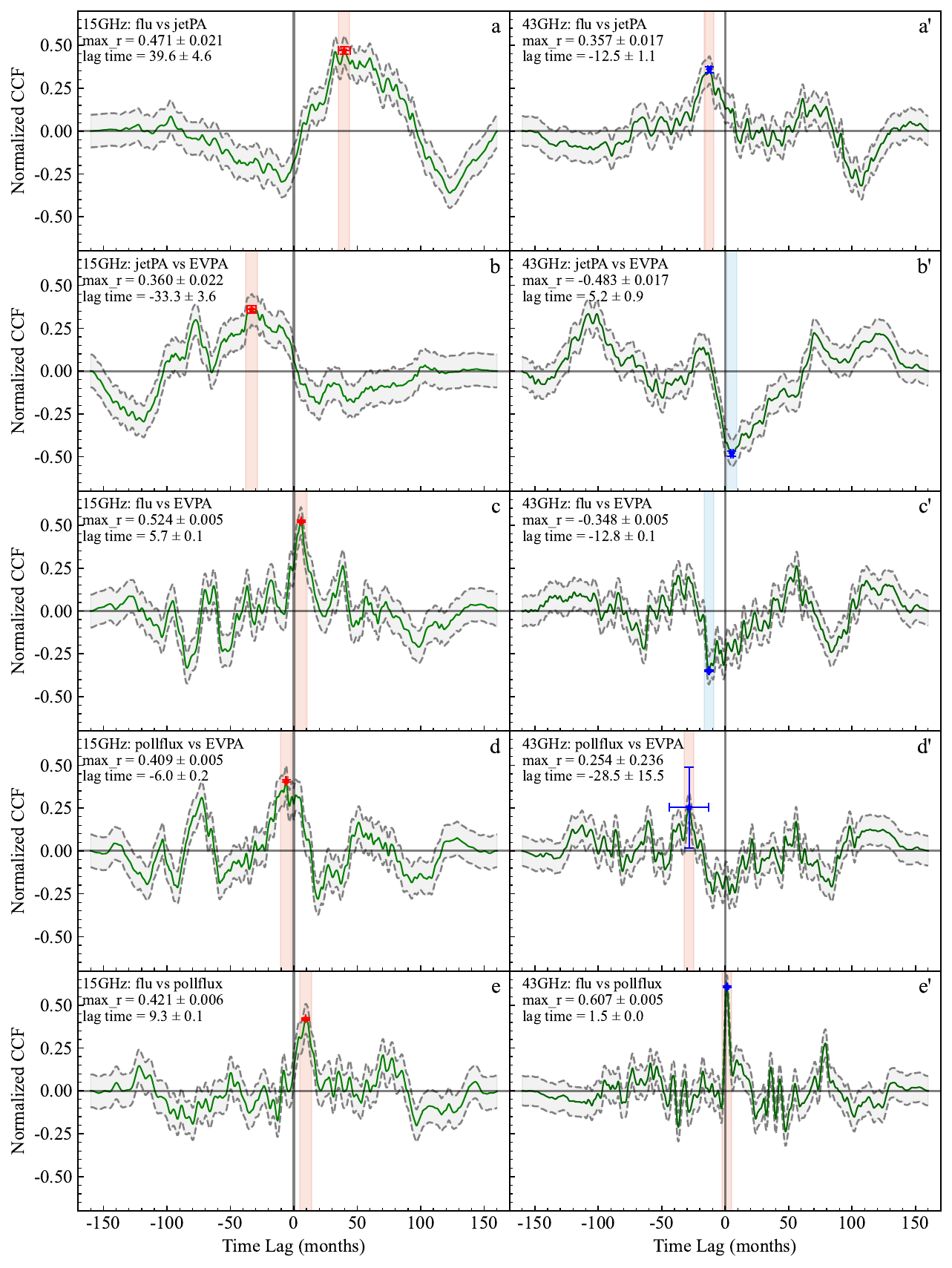}
\caption{Interpolation Cross-Correlation Function analyses of radio parameter pairs for the broad-line radio galaxy 3C 120, covering 15 GHz (left column) and 43 GHz (right column) bands. Panels from top to bottom for each frequency band: core flux vs jetPA, jetPA vs EVPA, core flux vs EVPA, pollflux vs EVPA, and core flux vs pollflux. Green solid curve: observed ICCF; gray curves: rigorous $3\sigma$ confidence envelopes from Monte Carlo simulations combining flux randomization and random subset selection. Red dots mark maximum correlation coefficient ($r$) and corresponding physical time lag. Colored vertical lines (pink for positive, light blue for negative correlation) indicate significant time lags.}
\label{fig:5}
\end{figure*}

To systematically evaluate the internal temporal couplings among the multi-frequency radio jet observables, we performed an ICCF analysis. We examined the pairwise relationships between core flux density, jet position angle, electric vector position angle, and polarized flux density. The resulting ICCF profiles for the 15~GHz and 43~GHz datasets are presented in Figure~\ref{fig:5}, and the corresponding peak correlation coefficients ($r$) and physical time lags ($\tau$) are summarized in Table~\ref{tab:internal_correlations}. Within our analytical framework, a positive time lag ($\tau > 0$) indicates that variations in the first parameter of each pair lead those in the second. We detected prominent cross-correlation peaks across multiple parameter pairs, revealing frequency-dependent physical linkages within the stratified jet.

For the core flux vs.\ jetPA correlations, we observe a frequency-dependent trend where changes in the jetPA lead the core-flux fluctuations at 43~GHz, whereas the opposite sequence is seen at 15~GHz. This behavior could be related to the stratified opacity structure of the jet, as the two frequencies probe different spatial scales due to synchrotron self-absorption. Specifically, at the compact, optically thick 43~GHz core, the emission traces the inner nozzle, where variations appear to be largely modulated by geometric and Doppler effects. As the macroscopic jet precession sweeps the nozzle orientation, the changing viewing angle can enhance the Doppler boosting factor, potentially amplifying the apparent emission and leading to a delayed core flux peak \citep{Villata1999, Caproni2004, Raiteri2017}. Conversely, at 15~GHz, which probes a more extended, optically thin downstream region, the kinematics may be more influenced by propagating perturbations. When localized energy injections (associated with core flux outbursts) travel along the pre-existing curved or helical jet trajectory, these active regions can sequentially illuminate different segments of the channel, introducing shifts in the flux-weighted centroid of the emission and contributing to a delayed variation in the measured jetPA \citep{Marscher2008}.

The correlations of EVPA with both jetPA and core flux provide a direct observational link between the structural morphology of the jet and its magnetic field topology. The diversity in correlation signs observed across different frequencies naturally reflects the complex 3D projection of the jet's intrinsic helical magnetic field. Physically, the emitting plasma is confined within a helical magnetic field bounded by the precessing jet cone \citep{Lyutikov2005, Marscher2008}. Because the inner radio core is spatially unresolved, the observed EVPA represents an integrated geometric projection. As a plasma parcel propagates, this projected EVPA is highly sensitive to the local azimuthal phase of the coiled magnetic field lines and the instantaneous viewing angle. Depending on this underlying, unresolved 3D geometry, the projected electric vectors can either track the bulk structural swing positively or rotate orthogonally \citep{Pushkarev2005, Cohen2015}. Consequently, these correlations suggest that the macroscopic structural morphology, total intensity, and magnetic field orientation are jointly modulated by a common geometric mechanism, such as secular jet precession \citep{Caproni2004, Raiteri2017}.

Beyond the geometric modulations imposed by the precessing nozzle, the intrinsic physical state of the jet plasma is encoded in the correlations among total core flux, polarized flux, and EVPA. We find consistent, strong positive dependencies for these parameter pairs at both 15~GHz and 43~GHz. The tight temporal coupling between total core flux and polarized flux indicates that macroscopic energy dissipation is intrinsically linked to the localized ordering of the magnetic field. As energetic perturbations evolve within the jet channel, they not only accelerate particles---driving the surges in total core flux---but also effectively realign the underlying turbulent magnetic field into a well-ordered topology. This rapid topological restructuring naturally produces the simultaneous surges in polarized emission \citep{Marscher2008}. Furthermore, the concurrent variation with EVPA indicates that this localized magnetic amplification is kinematically tied to the dynamic rotation of the polarization plane. Together, this coordinated behavior completes our unified physical picture: transient internal perturbations simultaneously govern both particle acceleration and the localized reconfiguration of the magnetic field geometry.

Synthesizing these multi-parameter correlations supports a unified dual-mechanism framework for the parsec-scale jet of 3C~120. The macroscopic structural orientation, long-term Doppler baseline evolution, and broad projected EVPA rotations are largely governed by secular jet precession. Because purely geometric models fail to account for the rapid, tightly coupled surges in both total and polarized fluxes, short-lived, energetic perturbations---such as internal shocks or magnetic reconnection events---are superimposed upon this slowly varying geometric baseline. These transient internal bursts account for the abrupt polarization surges and rapid magnetic topology reconfigurations observed during distinct flaring epochs.

\begin{deluxetable*}{lcccc}[ht]
\tablecaption{Internal Radio Correlation Results for 3C 120 at 15 GHz and 43 GHz \label{tab:internal_correlations}}
\tablewidth{0pt}
\tablehead{
\colhead{Parameter Pair} & \multicolumn{2}{c}{15 GHz} & \multicolumn{2}{c}{43 GHz} \\
\cline{2-3} \cline{4-5}
\colhead{} & \colhead{Correlation coefficient $r$} & \colhead{Time Lag $\tau$ (months)} & \colhead{Correlation coefficient $r$} & \colhead{Time Lag $\tau$ (months)}
}
\startdata
Core flux vs.\ Jet PA        & $0.471 \pm 0.021$ & $39.6 \pm 4.6$   & $0.357 \pm 0.017$  & $-12.5 \pm 1.1$  \\
Jet PA vs.\ EVPA             & $0.360 \pm 0.022$ & $-33.3 \pm 3.6$  & $-0.483 \pm 0.017$ & $5.2 \pm 0.9$    \\
Core flux vs.\ EVPA          & $0.524 \pm 0.005$ & $5.7 \pm 0.1$    & $-0.348 \pm 0.005$ & $-12.8 \pm 0.1$  \\
Polarized flux vs.\ EVPA     & $0.409 \pm 0.005$ & $-6.0 \pm 0.2$   & $0.254 \pm 0.236$  & $-28.5 \pm 15.5$ \\
Core flux vs.\ Polarized flux & $0.421 \pm 0.006$ & $9.3 \pm 0.1$    & $0.607 \pm 0.005$  & $1.5 \pm 0.0$    \\
\enddata
\tablecomments{Results of the ICCF analysis among internal radio observables at 15~GHz and 43~GHz. $r$ represents the peak cross-correlation coefficient, and $\tau$ represents the corresponding physical time lag. The stated uncertainties correspond to the 1$\sigma$ confidence intervals, derived from 10,000 monte carlo simulations combining flux randomization and gaussian peak-fitting. A positive time lag ($\tau > 0$) indicates that the variations in the first parameter lead those in the second parameter, whereas a negative lag indicates the reverse.}
\end{deluxetable*}

\subsection{Morphological Evolution and Superluminal Component Ejections}
\label{subsec:Morphological_Evolution}

\begin{figure*}[ht!]
\epsscale{1.0} 
\plotone{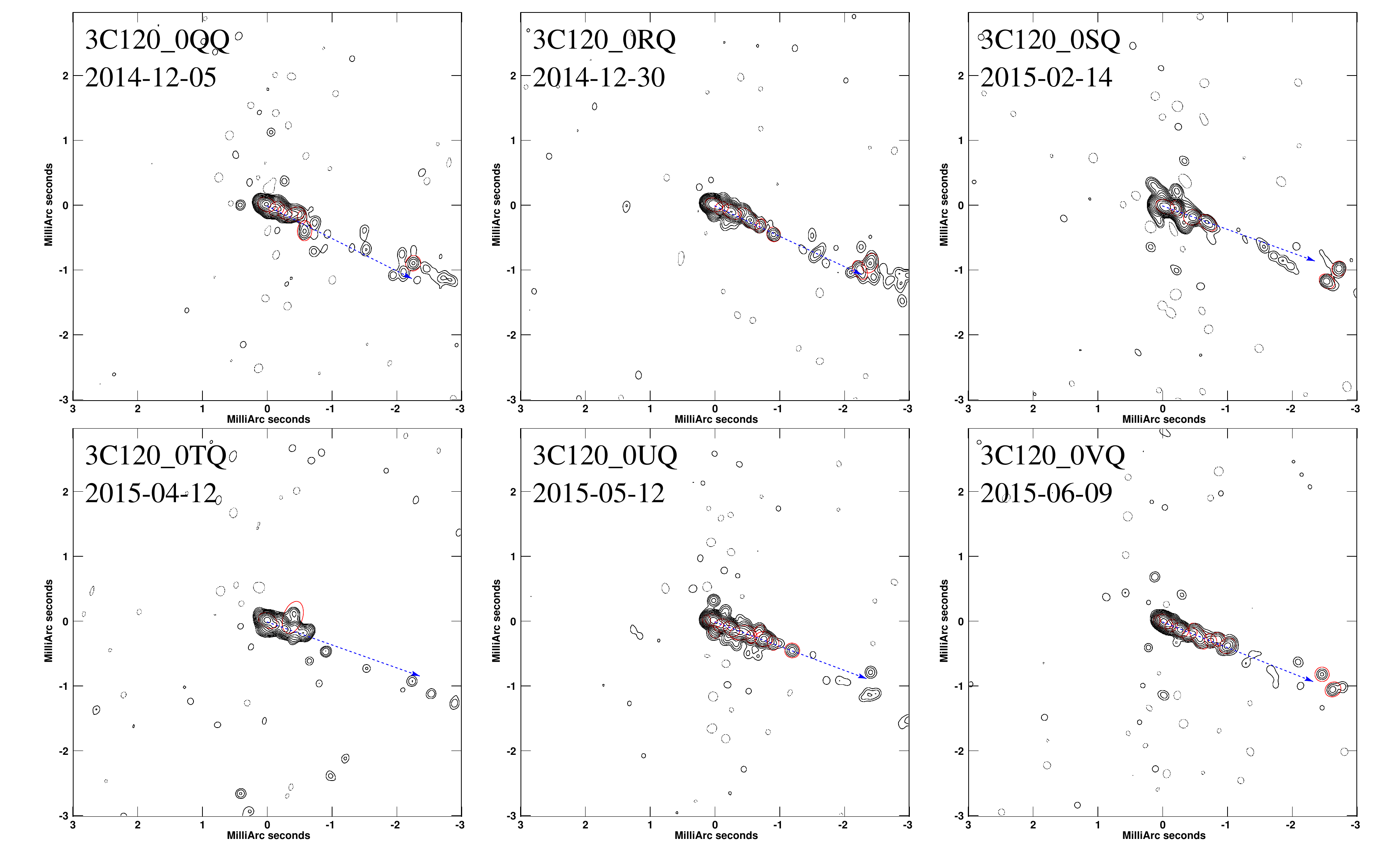}
\caption{Sequence of 43~GHz VLBA total intensity contour maps of 3C~120 during Flare I. The coordinate axes are in units of milliarcseconds (mas). Contours in the map are plotted at multiples of $-1, 1, 2, 4, 8, 16, 32, 64, 128, 256, 512, 1024 \times 4\sigma$, where $\sigma$ is the local RMS noise. The observation epochs in each panel are labeled using an extended HEX code, where the suffix ``Q'' denotes the Q-band (43~GHz).} 
\label{fig:8}
\end{figure*}

\begin{figure*}[ht!]
\epsscale{1.0} 
\plotone{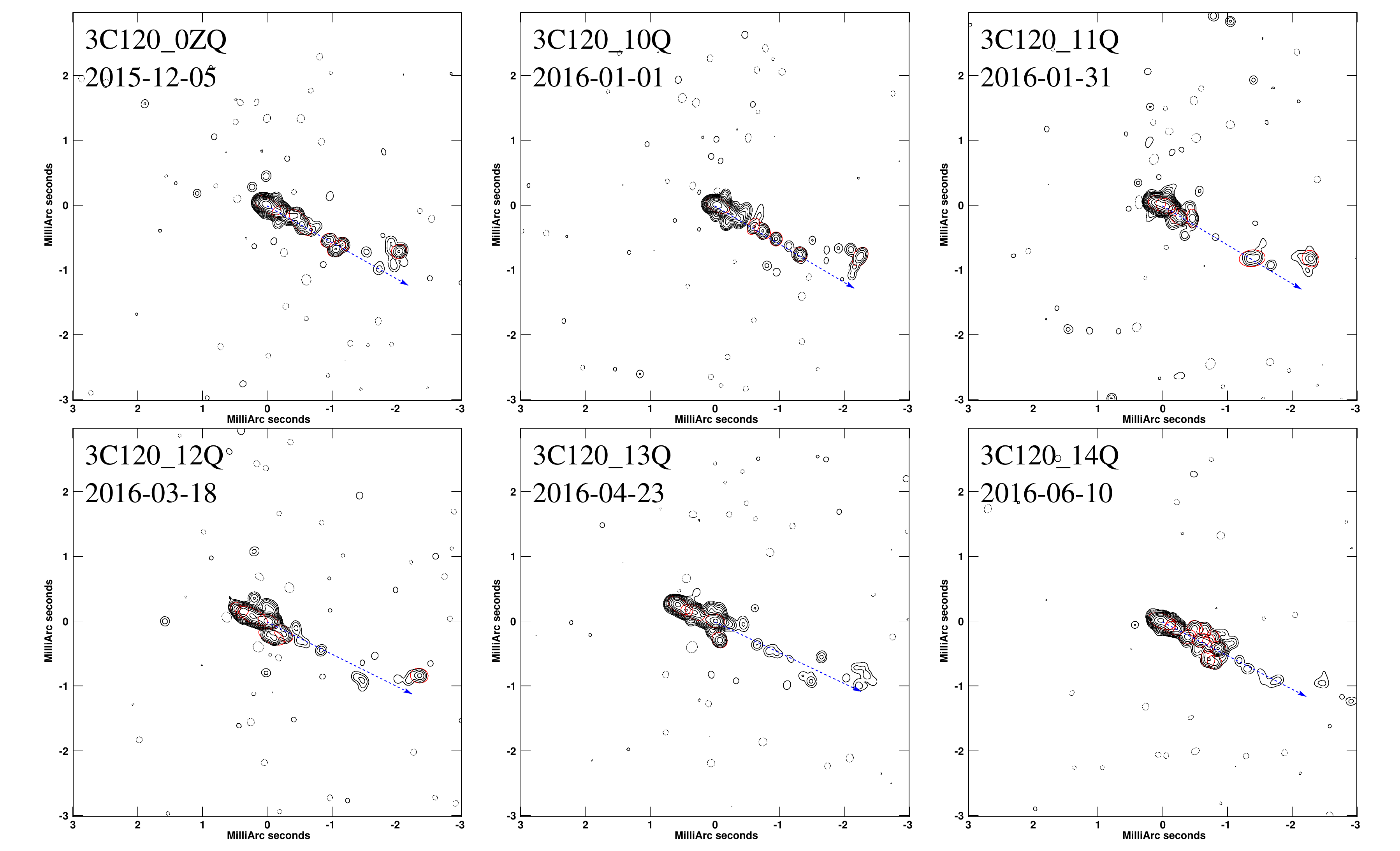}
\caption{Sequence of 43~GHz VLBA total intensity contour maps of 3C~120 during Flare II. The maps trace the morphological evolution of the parsec-scale jet across the flaring peak and subsequent decay phases. Contours and label descriptions are the same as those in Figures~\ref{fig:8}.}
\label{fig:9}
\end{figure*}

\begin{figure*}[ht!]
\epsscale{1.0} 
\plotone{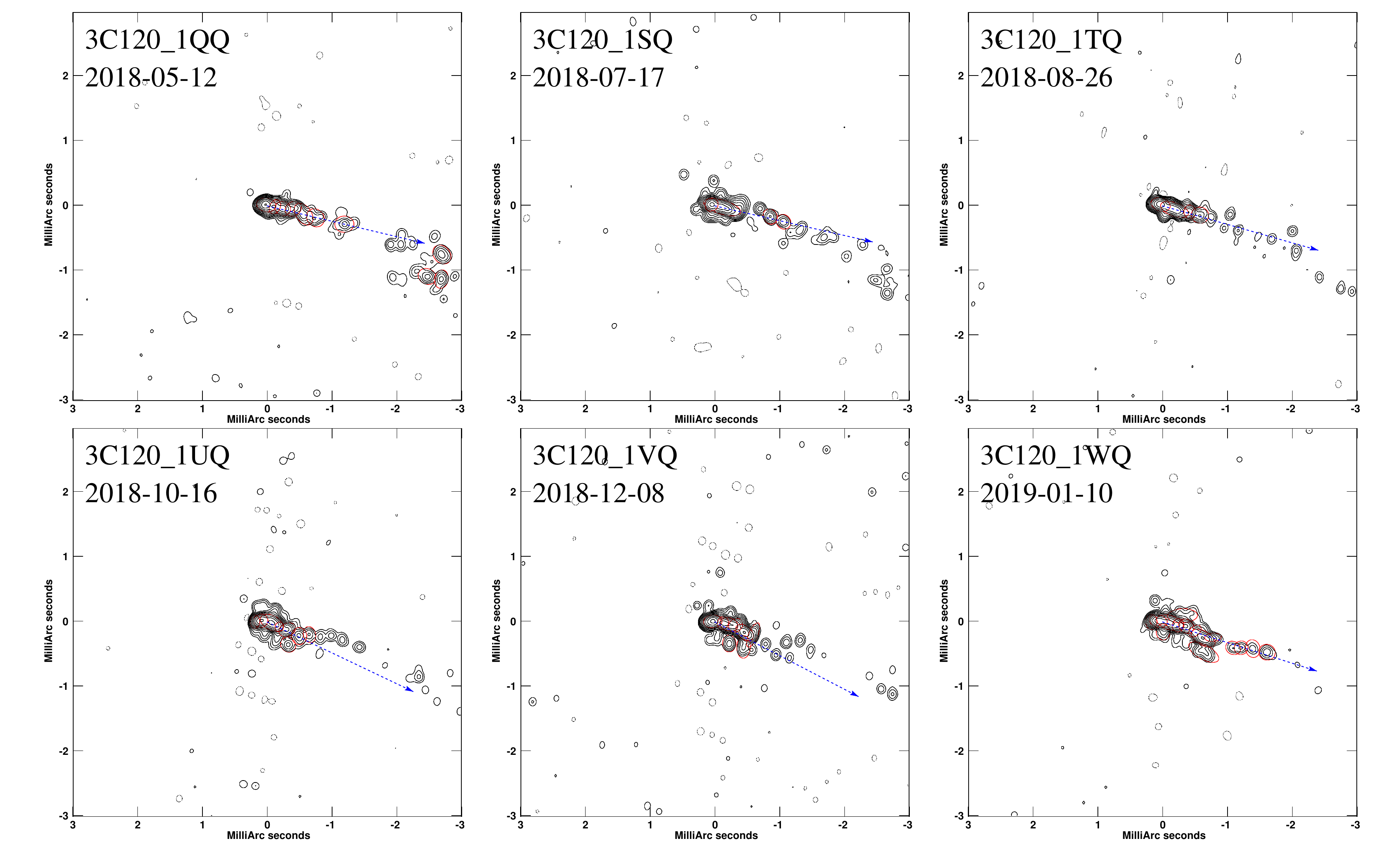}
\caption{Sequence of 43~GHz VLBA total intensity contour maps of 3C~120 during Flare III. The sequence captures the core region rapidly transitioning from a weak, diffuse structure to a highly compact, bright emission region peaking at 2.01~Jy at epoch 1UQ. Contours and label descriptions are the same as those in Figures~\ref{fig:8}.}
\label{fig:10}
\end{figure*}

To directly link the observed multi-wavelength flaring activities and core flux variations to the structural evolution of the 3C~120 parsec-scale jet, we analyzed the VLBA 43~GHz high-resolution total intensity contour maps from the BEAM-ME Program. Figures~\ref{fig:8}--\ref{fig:10} present the sequence of 43~GHz VLBA contour maps, focusing on the three major radio flaring epochs identified in Section~\ref{subsec:radio_flaring_polarization}. Additionally, Table~\ref{tab:2} summarizes the detailed observational data for these specific epochs, including the exact observing dates, core flux densities, and their corresponding uncertainties. Throughout all monitoring epochs, the large-scale jet of 3C~120 exhibits a consistent, stable extension toward the southwest. To visualize the jet kinematics, blue dashed arrows, the ODR linear fits of initial jet components, are overlaid on the contour maps to trace the trajectory of the bright and compact jet components  propagating downstream along the jet flow. The high-resolution morphological variations in the innermost jet region, when combined with the core flux and polarimetric evolution, provide direct imaging evidence for the dynamic processes driving the multi-wavelength outbursts.

During the Flare I, the jet structure remains stable from epoch 0QQ\footnote{The extended base-36 hexadecimal sequence number of the epoch in the SAND pipeline.} to 0RQ. Coinciding with the 43~GHz core flux rising from $\sim$1.52~Jy to a peak of 3.15~Jy at epoch 0SQ, the contour maps show dense, unresolved core emission. Subsequent epochs (0TQ, 0UQ, 0VQ) reveal a newly formed jet knot---explicitly marked by red circles and traced by blue dashed arrows---moving southwest. This confirms that unresolved core brightening precedes the ejection of a compact knot.

In Flare II, the core flux peaks at 4.00~Jy at epoch 12Q, appearing as an extremely bright, point-like source. The flux then drops sharply to $\sim$1.40~Jy by epoch 13Q, accompanied by the polarization surges reported in Section~\ref{subsec:radio_flaring_polarization}. Morphological maps from 13Q and 14Q show a distinct downstream elongation. Tracked by the red circles and blue arrows, this ``swelling'' represents an optically thin knot physically separating from the core, naturally explaining the rapid flux depletion.

In Flare III, the jet initially exhibits a relatively faint and extended core state across epochs 1SQ and 1TQ, with core fluxes of $\sim$0.77~Jy and $\sim$0.85~Jy, respectively. Subsequently, at epoch 1UQ, the core experiences a localized energy injection, brightening significantly to a highly compact peak of 2.01~Jy. Following this intense peak, the subsequent epochs (1VQ and 1WQ) capture a rapid depletion of the core flux, dropping to $\sim$0.63~Jy by 1WQ. Simultaneously, the contour maps reveal a pronounced downstream elongation and the emergence of a newly ejected knot. 

This morphological evidence is highly consistent with independent kinematic surveys \citep{Weaver2022}. In their analysis, newly ejected superluminal knots are identified as moving regions of enhanced radio emission within the parsec-scale jet. These discrete components are designated with a ``C'' prefix. Specifically, the observational sequence of Flare I coincides with the ejection and initial core-separation of knot C8. Similarly, the morphological sequence of Flare II aligns closely with the epoch range of the clustered ejections of subsequent components, namely C10 and C11. Furthermore, the sequence of Flare III temporally overlaps with the reported monitoring epochs of component C15.

Synthesizing these epochs, the structural evolution of 3C~120 strongly aligns with standard jet emission models (\citealt{Marscher1985}). All flares follow a consistent kinematic sequence: (1) optically thick core brightening via localized energy injection; (2) formation of a compact jet knot; (3) physical separation and downstream propagation, explicitly tracked by red circles and blue arrows; and (4) adiabatic expansion and fading. This high-resolution imaging confirms that internal shocks power the observed multi-wavelength flares.

\begin{deluxetable}{lcc}
\tablewidth{0pt}
\tablecaption{Observational Log of 43 GHz VLBI Monitoring during Major Flares}
\tablehead{
\colhead{Epoch} & \colhead{Time} & \colhead{Core Flux Density}
}
\startdata
\sidehead{\textbf{Flare I}}
3C120\_0QQ & 2014-12-05 & $1.524370 \pm 0.000471$ \\
3C120\_0RQ & 2014-12-30 & $1.660000 \pm 0.000000$ \\
3C120\_0SQ & 2015-02-14 & $3.149000 \pm 0.000000$ \\
3C120\_0TQ & 2015-04-12 & $2.220000 \pm 0.001000$ \\
3C120\_0UQ & 2015-05-12 & $1.406000 \pm 0.000000$ \\
3C120\_0VQ & 2015-06-09 & $2.203000 \pm 0.001000$ \\
\sidehead{\textbf{Flare II}}
3C120\_0ZQ & 2015-12-05 & $2.594000 \pm 0.000000$ \\
3C120\_10Q & 2016-01-01 & $3.078000 \pm 0.001000$ \\
3C120\_11Q & 2016-01-31 & $3.575000 \pm 0.001000$ \\
3C120\_12Q & 2016-03-18 & $4.000000 \pm 0.001000$ \\
3C120\_13Q & 2016-04-23 & $1.396130 \pm 0.000408$ \\
3C120\_14Q & 2016-06-10 & $1.220140 \pm 0.000634$ \\
\sidehead{\textbf{Flare III}}
3C120\_1QQ & 2018-05-12 & $1.15157 \pm 0.000411$ \\
3C120\_1SQ & 2018-07-17 & $0.770078 \pm 0.000333$ \\
3C120\_1TQ & 2018-08-26 & $0.848345 \pm 0.000353$ \\
3C120\_1UQ & 2018-10-16 & $2.01034 \pm 0.000521$ \\
3C120\_1VQ & 2018-12-08 & $1.31919 \pm 0.000391$ \\
3C120\_1WQ & 2019-01-10 & $0.632157 \pm 0.000323$ \\
\enddata
\label{tab:2}
\tablecomments{All data are derived from the 43~GHz BEAM-ME monitoring program. Columns: (1) Observation epoch ID; (2) Observation date; (3) Core flux density in Jy, with $1\sigma$ statistical uncertainties.}
\end{deluxetable}

\section{Discussion}
\label{sec:discussion}

\subsection{Physical Origins of the Internal Radio Couplings}
\label{subsec:precession_verification}

The ICCF results presented in Section~\ref{subsec:internal_radio_correlations} reveal a synchronized physical response among the internal radio observables. The strong coupling between the morphological parameter (jetPA) and the luminosity indicator (core flux), further corroborated by the correlation between jetPA and EVPA, indicates that these long-term variations are primarily driven by a slowly precessing jet cone. As the jet nozzle precesses, the varying viewing angle directly alters the Doppler boosting factor, thereby modulating the apparent core luminosity. Simultaneously, because the intrinsic magnetic field orientation is tightly ``locked'' to the geometric axis of the flow, this precessional motion largely governs the projected alignment of both jetPA and the magnetic field vector. Consequently, all physical observables bound to the precessing jet cone exhibit a synchronized ``co-movement'', jointly modulated by the changing line of sight.

However, while secular precession provides a consistent explanation for the long-term baseline trends, it cannot fully account for the rapid, dramatic flares observed in the polarized flux and the abrupt, short-timescale swings in EVPA (as highlighted in Figures~\ref{fig:3} and \ref{fig:4}). The characteristic timescales of these polarimetric bursts and magnetic field rotations are on the order of several months---drastically shorter than the $\sim 12.3$-year precession period previously established for 3C~120 \citep{Caproni2004}. Therefore, we argue that these rapid events are not artifacts of geometric viewing angle shifts. Instead, they represent a distinct, highly energetic internal mechanism. Analogous to a pulsar ``glitch''---where a steadily rotating lighthouse suddenly flickers due to internal magnetospheric instabilities---the 3C~120 jet undergoes steady global geometric precession that is occasionally punctuated by energetic, short-lived internal physical eruptions. These rapid events are likely driven by localized magnetic reconnection or the propagation of internal shocks that strongly compress the local magnetic field configuration, thereby triggering the observed intense, short-timescale polarization surges independent of the underlying slow precession.

\subsection{Multi-wavelength Variability and Kinematic Coupling}
\label{subsec:Optical_and_Radio_Kinematics}

While the cross-correlation between the $\gamma$-ray and radio bands yields prominent peaks, the global correlation between the detrended non-thermal optical residuals and the radio emission is statistically weak ($r_{\text{max}} \approx 0.14$--$0.27$). Physically, the multi-wavelength light curves are dominated by different mechanisms on global timescales. High-energy $\gamma$-ray and non-thermal optical emissions are produced deep in the upstream nozzle. Because these high-energy emitting electrons cool extremely rapidly, their light curves are predominantly driven by transient particle acceleration events, such as internal shocks or magnetic reconnection. Conversely, the parsec-scale radio emission is a complex superposition. While these upstream shocks eventually propagate downstream to trigger the short-lived radio bursts, the global variance of the radio light curve is heavily dominated by the long-term geometric modulation induced by secular jet precession. 

Consequently, a global cross-correlation analysis inherently mismatches the shock-dominated optical variations with the precession-dominated radio envelope, naturally diluting the statistical correlation strength \citep{Max-Moerbeck2014}. Furthermore, as individual shocks propagate downstream, they encounter differing initial conditions, jet opacities, and core-shift distances, introducing flare-dependent time delays that further smear the global correlation peak \citep{Chatterjee2009, Liodakis2018}.
However, this global dilution does not negate the presence of tightly coupled physical connections on localized timescales. Once the slowly varying geometric baseline is accounted for, an examination of the temporal data series confirms the presence of concurrent flux enhancements in the LOWESS-detrended optical residuals during these specific radio flaring periods. Furthermore, while our adaptive-binning methodology intrinsically smooths out weaker high-energy injections to maintain uniform statistical significance, the standard monthly-binned $\gamma$-ray data\footnote{\href{https://fermi.gsfc.nasa.gov/ssc/data/access/lat/LightCurveRepository/source.html?source_name=4FGL_J0433.0+0522}{Fermi-LAT Light Curve Repository: 4FGL J0433.0+0522}} verify the existence of corresponding minor flares that temporally align with these radio and optical activities.
This localized multi-band synchronicity robustly confirms that these radio outbursts are not isolated events. Instead, they represent the delayed downstream responses to the upstream internal shocks.

\section{Conclusions}
\label{sec:Conclusions}

In this paper, we present a comprehensive 13-year (2012--2025) multi-wavelength study of the broad-line radio galaxy 3C~120, combining \textit{Fermi}-LAT, ASAS-SN, and high-resolution VLBA (15~GHz and 43~GHz) observations. By systematically analyzing the time-series data and multi-epoch morphological jet tracking, we dynamically link the high-energy dissipation microphysics to macroscopic parsec-scale jet dynamics. Our cross-correlation analyses reveal a clear frequency-dependent temporal sequence: $\gamma$-ray variations systematically lead the 43~GHz and 15~GHz radio emission by $8.27_{-5.55}^{+3.45}$ and $11.08_{-1.88}^{+4.03}$ months, respectively. This frequency hierarchy confirms that the primary high-energy dissipation zone is located at parsec scales upstream of the radio core, providing robust dynamical support for the opacity-driven core-shift effect.

High-resolution 43~GHz VLBA imaging clearly captures the downstream responses to these high-energy triggers. Consistent with the canonical shock-in-jet model, the major flaring epochs are accompanied by unresolved core brightening, dramatic polarization surges, rapid EVPA rotations, and the subsequent ejection of superluminal knots. 
To explain the diverse multi-wavelength behaviors and structural evolution of 3C~120, we propose a dual-mechanism framework. The optical and $\gamma$-ray variations are directly driven by transient particle acceleration events, such as internal shocks. When these shocks propagate downstream into the radio regions, their signatures are superimposed onto a slowly varying geometric baseline. Consequently, secular jet precession governs the long-term radio flux and macroscopic structural swings, while episodic internal shocks drive the rapid polarimetric bursts and localized radio flares.

Ultimately, this multi-wavelength synthesis yields important insights into general AGN jet physics, which is potentially applicable to the broader blazar population. Future multi-messenger and high-resolution observations (e.g., ngVLA, CTA, IceCube-Gen2) will be crucial to rigorously constrain the fine spatial structure of the jet dissipation zone and identify the ultimate physical drivers of relativistic jet activity.

\begin{acknowledgments}
This work was supported by the National Key R\&D Program of China (Intergovernmental Cooperation, 2024YFA1611500) and the National Natural Science Foundation of China (12173078).
This research has made use of data from the MOJAVE database that is maintained by the MOJAVE team \citep{Lister2018}.
This study makes use of VLBA data from the VLBA-BU Blazar Monitoring Program (BEAM-ME and VLBA-BU-BLAZAR; \href{http://www.bu.edu/blazars/BEAM-ME.html}{http://www.bu.edu/blazars/BEAM-ME.html}), funded by NASA through the Fermi Guest Investigator Program. The VLBA is an instrument of the National Radio Astronomy Observatory. The National Radio Astronomy Observatory is a facility of the National Science Foundation operated by Associated Universities, Inc.
This research has made use of data from the ASAS-SN project, which is supported by the Ohio State University and operated by the Ohio State Astronomy Department.

Software: AIPS \citep{Greisen2003}, Python, SAND \citep{zhang2016}.
\end{acknowledgments}

\bibliography{sample701}{}
\bibliographystyle{aasjournalv7}

\end{document}